\documentclass[lettersize,journal]{IEEEtran}

\usepackage[T1]{fontenc}
\usepackage[utf8]{inputenc}

\usepackage{amsmath,amsfonts}
\usepackage{algorithmic}
\usepackage{array}
\usepackage[caption=false,font=normalsize,labelfont=sf,textfont=sf]{subfig}
\usepackage{textcomp}
\usepackage{stfloats}
\usepackage{url}
\usepackage{verbatim}
\usepackage{graphicx}
\usepackage{cite}
\usepackage{multirow}
\usepackage{booktabs}
\usepackage[hidelinks]{hyperref}

\usepackage[acronym]{glossaries}
\makenoidxglossaries
\setacronymstyle{long-short}

\DeclareUnicodeCharacter{2212}{\ensuremath{-}} 
\DeclareUnicodeCharacter{2013}{--}            
\DeclareUnicodeCharacter{2014}{---}           

\newacronym{ai}{AI}{Artificial Intelligence}
\newacronym{artex}{ARTEX}{Army Technological Exercise}
\newacronym{astar}{A*}{A-star}
\newacronym{ba}{BA*}{Bidirectional A*}
\newacronym{bfs}{BFS}{Breadth-First Search}
\newacronym{c2}{C2}{Command and Control}
\newacronym{c4isr}{C4ISR}{Command, Control, Communications, Computers, Intelligence, Surveillance and Reconnaissance}
\newacronym{cos}{COS}{Carta de Uso e Ocupação do Solo}
\newacronym{cpa}{CPA}{Closest Point of Approach}
\newacronym{dem}{DEM}{Digital Elevation Model}
\newacronym{dijkstra}{Dijkstra's}{Dijkstra's Algorithm}
\newacronym{dimacs}{DIMACS}{Center for Discrete Mathematics and Theoretical Computer Science}
\newacronym{dstar}{D*}{D-star}
\newacronym{gis}{GIS}{Geographic Information System}
\newacronym{gnss}{GNSS}{Global Navigation Satellite System}
\newacronym{gsm}{GSM}{Global System for Mobile Communications}
\newacronym{isr}{ISR}{Intelligence, Surveillance, and Reconnaissance}
\newacronym{kpi}{KPI}{Key Performance Indicator}
\newacronym{lpa}{LPA*}{Lifelong Planning A*}
\newacronym{nato}{NATO}{North Atlantic Treaty Organization}
\newacronym{rcspp}{RCSPP}{Resource-Constrained Shortest Path Problem}
\newacronym{repmus}{REPMUS}{Robotic Experimentation and Prototyping using Maritime Unmanned Systems}
\newacronym{snr}{SNR}{Signal-to-Noise Ratio}
\newacronym{ugv}{UGV}{Unmanned Ground Vehicle}
\newacronym{wc-a}{WC-A*}{Weight--Constrained A*}
\newacronym{wc-ba}{WC-BA*}{Weight--Constrained Bidirectional A*}
\newacronym{wc-ebba}{WC-EBBA*}{Weight--Constrained Enhanced Bidirectional Bounded A*}
\newacronym{argus}{ARGUS}{Adaptive Route Guidance and Update System}
\newacronym{apulse}{APULSE}{A*-Pulse}

\begin{document}

\title{ARGUS: A Framework for Risk-Aware Path Planning in Tactical UGV Operations}

\author{
    \begin{tabular}[t]{c} 
        Nuno Alexandre Duarte Soares \\
        \textit{Academia Militar} \\
        Lisboa, Portugal \\
        soares.nad@academiamilitar.pt
    \end{tabular}
    \hspace{2cm} 
    \begin{tabular}[t]{c} 
        António Manuel Raminhos Cordeiro Grilo \\
        \textit{INESC INOV} \\
        \textit{Instituto Superior Técnico (IST)} \\
        \textit{Universidade de Lisboa} \\
        Lisboa, Portugal \\
        antonio.grilo@inov.pt
    \end{tabular}
}

\maketitle

\begin{abstract}
This thesis presents the development of ARGUS, a framework for mission planning for Unmanned Ground Vehicles (UGVs) in tactical environments. The system is designed to translate battlefield complexity and the commander's intent into executable action plans. To this end, ARGUS employs a processing pipeline that takes as input geospatial terrain data, military intelligence on existing threats and their probable locations, and mission priorities defined by the commander. Through a set of integrated modules, the framework processes this information to generate optimized trajectories that balance mission objectives against the risks posed by threats and terrain characteristics. A fundamental capability of ARGUS is its dynamic nature, which allows it to adapt plans in real-time in response to unforeseen events, reflecting the fluid nature of the modern battlefield. The system's interoperability were validated in a practical exercise with the Portuguese Army, where it was successfully demonstrated that the routes generated by the model can be integrated and utilized by UGV control systems. The result is a decision support tool that not only produces an optimal trajectory but also provides the necessary insights for its execution, thereby contributing to greater effectiveness and safety in the employment of autonomous ground systems.
\end{abstract}


\section{Introduction}
\IEEEPARstart{T}{he} character of modern warfare has shifted significantly. Conflicts are no longer defined solely by the massing of troops or superiority in firepower, but increasingly by the ability to operate effectively in sensor-saturated environments where every movement can be detected, tracked, and targeted. In such settings, survivability depends on minimizing exposure, maintaining mobility, and reacting rapidly to evolving threats. \glspl{ugv} have emerged as a transformative capability in this new operational reality~\cite{salacinski2019unmanned}. They provide commanders with the means to extend their operational reach, execute missions in dangerous areas, and enhance decision-making through richer, faster situational awareness. By reducing the need to expose soldiers to direct danger, \glspl{ugv} are fundamentally changing ground operations.

This paradigm shift is reflected in the strategic priorities of modern armed forces. The Portuguese Army, for instance, is making a significant investment in autonomous systems—an initiative that embodies the urgent need to intelligently integrate \glspl{ugv} as core components of the future force~\cite{Candido2025Lessons}. This national priority is reinforced at the Alliance level, where \gls{nato} has signaled that \gls{ai}-enabled autonomy and unmanned systems will be decisive across \gls{c4isr}, logistics, and manoeuvre. The \gls{nato} Science and Technology Organization (STO) projects a revolutionary impact of \gls{ai} and autonomy on military operations out to 2040~\cite{STO2020Trends}, underscoring the timeliness of rigorous research into planning methods that make \gls{ugv} employment safer, faster, and more reliable.

Despite their promise, current \gls{ugv} deployments reveal a critical capability gap. Most systems remain task-specific, with a limited ability to adapt dynamically to a changing operational environment. A logistics \gls{ugv} may follow a pre-planned path efficiently but be incapable of adjusting its route in response to a sudden threat; a reconnaissance platform may gather valuable data but lack the autonomy to intelligently avoid detection. This lack of adaptability undermines their value in complex, real-world missions. Military doctrine has already recognized this shortfall, with \gls{nato} stressing the need for resilient, interoperable, and risk-aware autonomous systems whose effective integration is critical to achieving operational superiority~\cite{nato2011capability,nato2016ajp3}. The challenge, therefore, is not merely to build a vehicle that moves autonomously, but to enable that vehicle to move intelligently—navigating while accounting for threats, minimizing detection, and balancing mission objectives. This transforms path planning from a technical problem into a doctrinal imperative.

Recent military experimentation, such as the exercises conducted by the Portuguese Army \gls{artex} and Navy \gls{repmus}, provides valuable insight into this challenge. While demonstrating the growing maturity of \gls{ugv} platforms, these exercises highlighted that dynamic adaptability remains a decisive factor for mission success~\cite{Candido2025Lessons}. In these operationally complex environments, static plans are insufficient; the ability to perform rapid, local replanning is a prerequisite for both survivability and mission success. However, the academic field still lacks models that adequately capture these operational requirements. Existing approaches often focus on vehicle autonomy at a technical level but provide limited support to the high-level decision-making processes that must guide \gls{ugv} movement across the battlefield. As noted in recent work, there is an absence of established methods for tactical terrain analysis that directly support autonomous \gls{ugv} mission planning and commander decision-making~\cite{Maaiveld2024}.

\begin{figure}[!t]
    \centering
    \includegraphics[width=0.95\columnwidth]{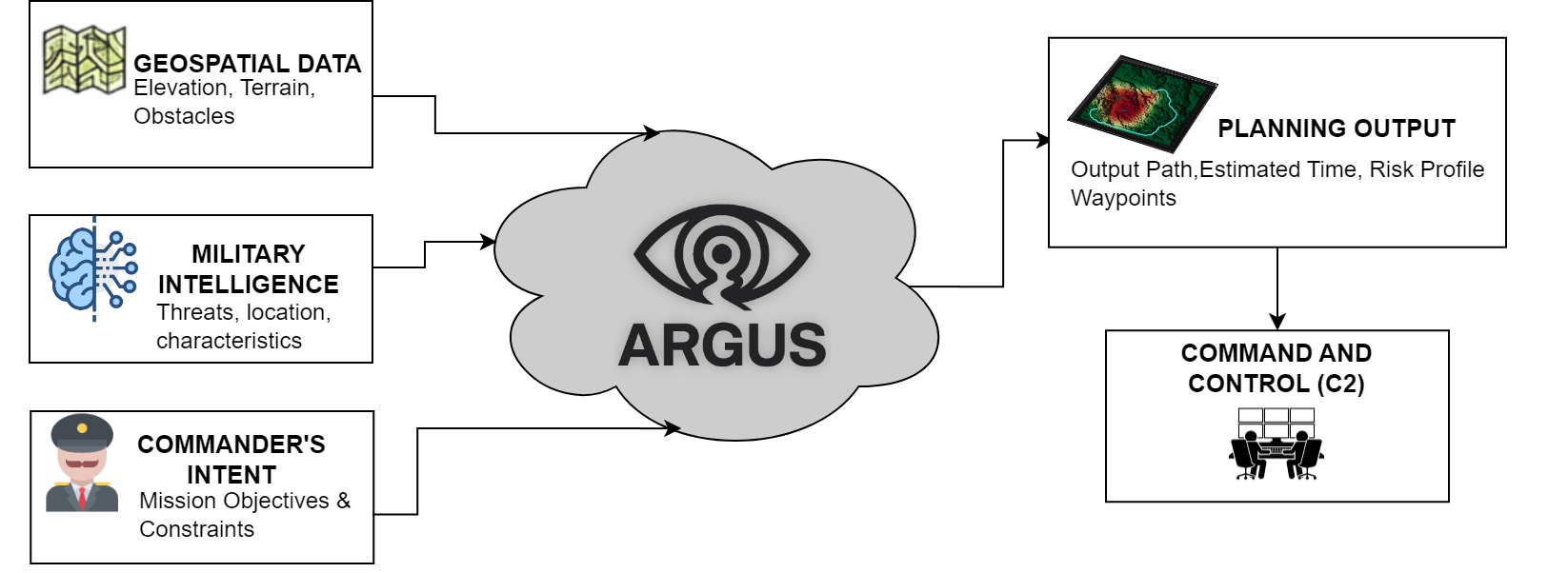}
    \caption{Conceptual diagram of the ARGUS planning framework, showing the integration of geospatial data, military intelligence, and the commander’s intent to produce actionable, risk-aware paths for \glspl{ugv}.}
    \label{fig:argus_conceptual_overview}
\end{figure}

What is required is not another navigation algorithm, but a mission-centric planning framework that can translate a commander’s intent, the dynamics of the tactical situation, and the inherent trade-offs between time and risk into an executable course of action. To address this challenge, this thesis presents \gls{argus}, a risk-aware, multi-objective path planning framework designed to meet the demands of modern military operations. As illustrated in Fig.~\ref{fig:argus_conceptual_overview}, \gls{argus} integrates three fundamental sources of information: geospatial data (elevation, terrain, obstacles), military intelligence (threats, locations, characteristics), and the commander’s intent (objectives and constraints). The framework processes these inputs to generate optimized paths that reflect both the operational environment and the commander's planning guidance, closing the loop between mission planning and execution.

This thesis makes several key contributions to address the identified capability gap. First, a unified risk modeling framework is introduced that integrates geospatial data, probabilistic threat intelligence, and the \gls{ugv} formation’s physical footprint to generate a tactically relevant cost surface. Second, a suite of three mission-driven planning modes is presented, which formalize a commander’s intent into distinct optimization problems, enabling flexible adaptation to diverse tactical priorities. Third, a proprietary \gls{apulse} algorithm, developed as a highly efficient solver for the time-constrained, risk-minimization problem, a form of the \gls{rcspp}, demonstrating superior scalability on large, real-world graphs. Finally, the results that include the mission paths generated by the framework and the comparative performance analysis of the developed algorithm against state-of-the-art competitors.

The remainder of this thesis is organized as follows. Section~II reviews the state of the art in path planning and risk modeling. Section~III details the methodology of the framework components. Section~IV describes the \gls{apulse} algorithm. Section~V presents the prototype implementation and operational demonstration. Section~VI provides a comprehensive evaluation of the framework and algorithm performance. Finally, Section~VII concludes the thesis and discusses future research directions.

\section{Related Work}
\noindent The development of a mission-centric planning framework for \glspl{ugv} builds upon established research in world representation, graph search algorithms, constrained optimization, and risk modeling. This section reviews the foundational concepts in these areas to contextualize the contributions of this thesis.

\subsection{Graph-Based World Representation and Pathfinding}
The foundational task in path planning is the translation of a complex, continuous operational environment into a discrete, machine-readable format~\cite{LaValle2006}. The state-of-the-art methodology represents the world as a weighted graph, $G=(V, E)$, where pathfinding algorithms can be executed~\cite{Ferguson2007}. The process typically begins by discretizing the terrain into a regular grid, where each cell becomes a node in the graph. To imbue this structure with meaning, it is enriched with multiple layers of geospatial data. A \gls{dem} is critical for providing altitude data, from which traversal metrics like terrain slope can be derived~\cite{Wang2024}. Additional layers, such as land-cover data, are integrated to classify surface types (e.g., forest, asphalt), which heavily influence vehicle speed and mobility~\cite{Vandapel2004ICRA}. The final step is to assign a non-negative, additive cost to each edge, representing a quantitative measure of traversability based on these factors.

Given this weighted graph, pathfinding becomes a classical shortest path problem. The evolution of algorithms to solve this problem has progressed from uninformed to informed, goal-directed methods. \gls{dijkstra} algorithm serves as the seminal, uninformed benchmark; it guarantees optimality by exploring the graph in expanding waves of equal cost but is computationally inefficient for large maps due to its exhaustive nature~\cite{Dijkstra1959}. A decisive step forward was the \gls{astar} algorithm, which introduced heuristic guidance to the search process~\cite{Hart1968}. By combining the actual cost from the start with an estimated cost to the goal, \gls{astar} prioritizes promising directions, drastically reducing the number of explored nodes while maintaining optimality guarantees. Further enhancements led to bidirectional variants like \gls{ba}, which conduct two simultaneous searches from the start and goal nodes. This approach has been consistently shown to reduce computational effort, particularly in large-scale pathfinding problems, by exploring two smaller search spaces that meet in the middle~\cite{symmetry2021}~\cite{Holte2016}.

To illustrate the performance difference between uninformed and informed algorithms, Table~\ref{tab:maze_runner} summarizes a benchmark comparison of \gls{astar}, \gls{dijkstra}, and \gls{bfs} in a maze-navigation scenario~\cite{Permana2018}. Although all methods reached identical optimal paths, \gls{astar} achieved a significant reduction in the number of explored nodes, highlighting the computational advantages of heuristic guidance.

\begin{table}[!t]
\centering
\caption{Comparison of \gls{astar}, \gls{dijkstra}, and \gls{bfs} performance on the Maze Runner Game (adapted from~\cite{Permana2018}).}
\label{tab:maze_runner}
\resizebox{\columnwidth}{!}{
\begin{tabular}{lccc}
\toprule
Component & A* Algorithm & Dijkstra & BFS \\
\midrule
Path length & 38 & 38 & 38 \\
Computation time (ms) & 0.35 & 0.30 & 0.80 \\
Nodes expanded & 323 & 738 & 738 \\
\bottomrule
\end{tabular}}
\end{table}

\subsection{The Resource-Constrained Shortest Path Problem (RCSPP)}
Military missions are rarely single-objective. The need to find the safest path that adheres to a strict time budget is a classic \gls{rcspp}, a problem known to be NP-hard. This complexity renders standard shortest-path algorithms insufficient. Historically, exact solutions based on dynamic programming and label-correcting algorithms have been proposed~\cite{Lozano2013}. However, these methods suffer from a “curse of dimensionality,” where the number of potentially optimal (non-dominated) labels at each node can grow exponentially with the size of the graph, leading to prohibitive memory consumption and computation times on large networks~\cite{Pugliese2013}.

To overcome these scalability limitations, the modern paradigm for solving the \gls{rcspp} has converged on sophisticated label-setting algorithms that synergistically combine three key mechanisms. First, they employ an informed, best-first search strategy, with \gls{astar} and its variants serving as the predominant foundation~\cite{Thomas2018}. Second, they incorporate aggressive real-time pruning techniques, often inspired by Pulse algorithms, which use strong feasibility and optimality checks to dynamically discard large, unpromising portions of the search tree~\cite{Cabrera2020}. Third, to control label proliferation, they use state-space reduction techniques, most notably the discretization of resource dimensions into “buckets,” which trades a controlled loss of precision for substantial improvements in performance~\cite{Delling2009}~\cite{Irnich2005}. Benchmark comparisons on large-scale road networks, such as the \gls{dimacs} instances, confirm that hybrid algorithms integrating these three mechanisms achieve runtimes that are orders of magnitude faster than classical or single-strategy approaches~\cite{Ahmadi2021}.

The performance of representative algorithms on \gls{dimacs} road-network instances is summarized in Table~\ref{tab:rcspp_benchmark}~\cite{Ahmadi2021}. The results demonstrate the superior scalability of recent label-setting and bidirectional heuristic approaches such as RC-EBBA*, which inspire the computational strategy adopted by the APULSE algorithm.

\begin{table}[!t]
\centering
\caption{Average runtime and number of solved instances for major \gls{rcspp} algorithms on \gls{dimacs} road networks (adapted from~\cite{Ahmadi2021}).}
\label{tab:rcspp_benchmark}
\resizebox{\columnwidth}{!}{
\begin{tabular}{lrrrrr}
\toprule
Instance & Nodes (×10\textsuperscript{6}) & K-SP & Pulse & RC-BDA* & RC-EBBA* \\
\midrule
NY   & 0.26 & 0.42 s & 15.54 s & 0.58 s & 0.07 s \\
BAY  & 0.32 & 2.36 s & 14.86 s & 3.94 s & 0.08 s \\
COL  & 0.44 & 7.73 s & 1446.77 s & 15.07 s & 0.11 s \\
FLA  & 1.07 & 60.80 s & 3743.91 s & 33.43 s & 0.42 s \\
NE   & 1.52 & 1.98 s & 978.92 s & 2.26 s & 0.12 s \\
CAL  & 1.89 & 2430.94 s & 4330.04 s & 3145.20 s & 11.49 s \\
LKS  & 2.76 & 1354.91 s & 5761.24 s & 1840.40 s & 6.24 s \\
E    & 3.60 & 46.88 s & 3652.11 s & 739.26 s & 0.69 s \\
W    & 6.26 & 858.04 s & 7479.83 s & 3410.24 s & 3.00 s \\
CTR  & 14.08 & 1155.85 s & 8051.86 s & 981.45 s & 5.51 s \\
USA  & 23.95 & 5009.80 s & 8645.88 s & 6293.76 s & 89.99 s \\
\midrule
Overall avg. & — & 993.61 s & 4010.99 s & 1496.87 s & 10.70 s \\
\bottomrule
\end{tabular}}
\end{table}

\subsection{Dynamic Replanning and Incremental Search}
Classical pathfinding algorithms assume a static environment. In real-world operations, however, new information about threats or obstacles can invalidate a pre-computed path. The most direct response, a full replan from scratch, is profoundly inefficient as it re-computes large amounts of information that remain unchanged~\cite{stentz1994optimal}. This computational redundancy is unsuitable for applications requiring rapid decision-making.

To address this, the paradigm of incremental search was developed. Algorithms in this class are designed to efficiently repair an existing solution by reusing information from previous computations, focusing effort only on the regions of the graph affected by the new information. The \gls{dstar} algorithm and its more efficient successor, \gls{dstar} Lite, are pioneering works in this area~\cite{koenig2002d}~\cite{stentz1995d}. \gls{dstar} Lite, in particular, has become a foundational technique in robotics. When an edge cost changes, it does not invalidate the entire search tree but instead locally propagates cost updates from the point of change until the path is optimal again. This path-repair approach provides correctness guarantees with significantly less computation than a full \gls{astar} search, making it a suitable foundation for the dynamic adaptation required in military operations.

\subsection{Detection-Probability Modeling for Risk Assessment}
A credible path planner must be grounded in a principled model of detection risk. The analytical foundations for this are provided by radar detection theory, which establishes that the probability of detection ($P_d$) follows a characteristic sigmoidal (S-shaped) curve as a function of range~\cite{Khawaja2022}. This behavior arises from the relationship between the signal-to-noise ratio \gls{snr} and $P_d$ within the Neyman–Pearson framework~\cite{Marcum1948}~\cite{Swerling1960}.

While rigorous physical models like the Marcum Q-function are computationally prohibitive for real-time planning, the standard practice in mission planning is to adopt computationally cheaper, closed-form parametric approximations that replicate the characteristic sigmoidal plateau-and-decay profile~\cite{Mahafza2003}~\cite{Kabamba2006}. The validity of this approach has been demonstrated in recent work, which shows that the parameters of these simplified models can be grounded in the underlying physics or calibrated from empirical data, ensuring they remain a faithful and interpretable representation of the real detection phenomenon~\cite{Costley2022Sensitivity}. Furthermore, this methodology has been successfully validated in challenging, non-ideal scenarios, such as passive radar systems, confirming its practical utility for real-world mission planning~\cite{Droge2022, Peto2013}.

Despite this progress, the literature review identified two critical gaps. First, there is a lack of a consistent, unified model that translates detection theory into a quantifiable, additive risk cost suitable for graph search with \glspl{ugv}. Second, existing work does not adequately address the quantification of detection risk for ground vehicles with a non-zero physical footprint, such as a \gls{ugv} formation. This thesis directly addresses these gaps.

\section{The ARGUS Methodology}
\noindent The \gls{argus} framework is a modular, layered system designed to transform heterogeneous inputs—geospatial data, military intelligence, and the commander's intent—into an optimized and executable path for a \gls{ugv}. This section details the methodology underpinning the framework, from data ingestion and world modeling to probabilistic risk formulation and mission-driven optimization.

\subsection{System Architecture and Data Flow}
\begin{figure}[!t]
    \centering
    \includegraphics[width=0.9\columnwidth]{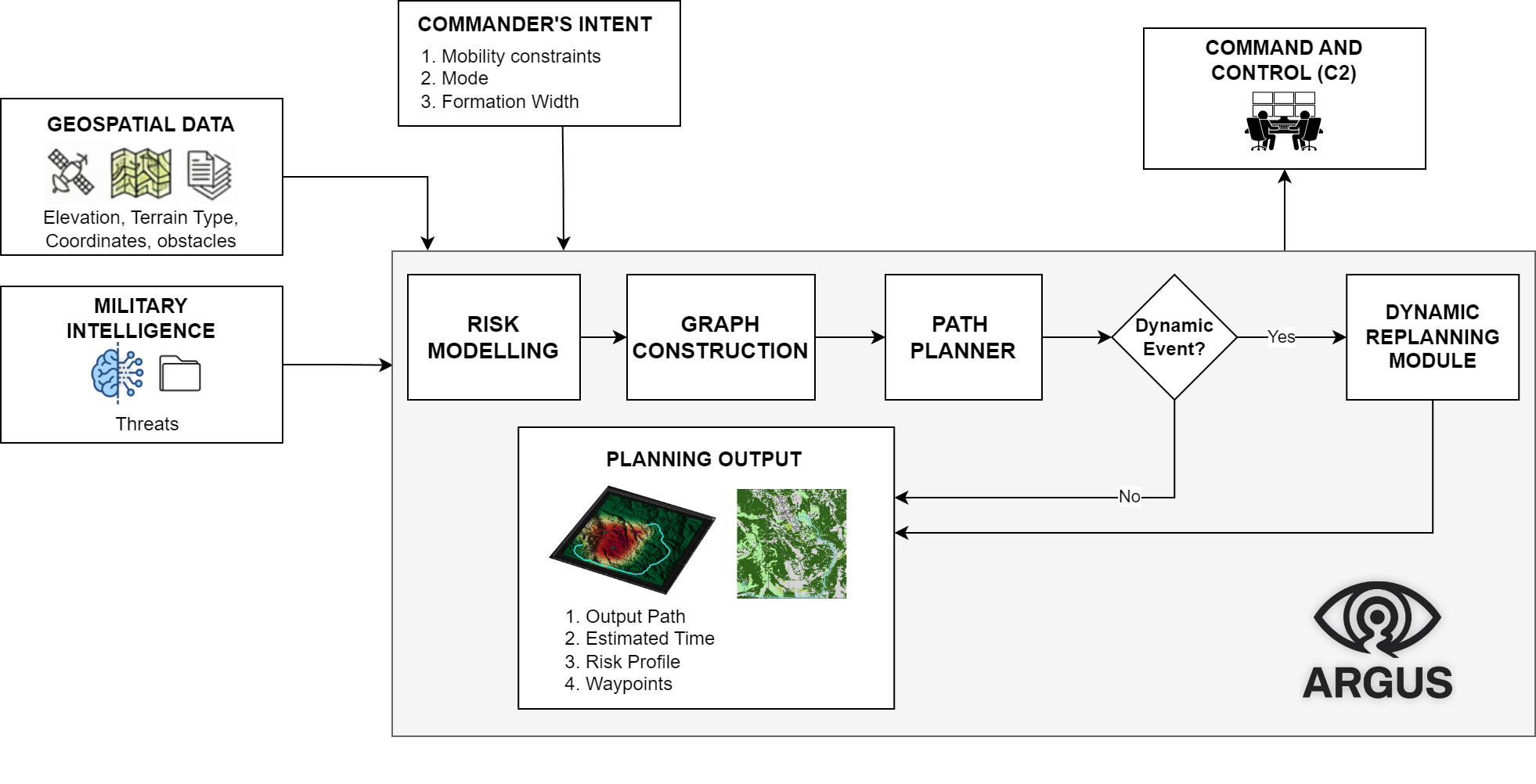}
    \caption{ARGUS system architecture and data flow from ingestion to mission-ready path. The modular design separates data processing, risk modeling, and planning into distinct interoperable stages.}
    \label{fig:argus_architecture}
\end{figure}

The \gls{argus} processing pipeline, shown in Fig.~\ref{fig:argus_architecture}, is designed for modularity and traceability. The workflow begins with terrain pre-processing, where geospatial layers are harmonized and mobility-related attributes, such as slope, are derived. In parallel, the model is designed to ingest data provided by intelligence teams, which associate each detected or inferred threat with a per-cell probability of location and a corresponding set of threat characteristics, thus forming a spatially distributed locational prior.

These two streams converge in the risk evaluation stage, which synthesizes a per-cell risk surface. This surface is then used to construct a multi-cost weighted graph serving as the foundation for planning. The model computes an optimal path according to the commander-defined mission mode and operational constraints. The architecture also includes a dynamic replanning module that locally repairs paths when new information becomes available, avoiding full recomputation. This modular design enables independent tuning of components without affecting global interoperability.

\subsection{World Modeling and Graph Construction}
The operational environment is modeled as a weighted graph $G=(V,E)$, where $V$ denotes the set of nodes (terrain cells) and $E$ the set of feasible edges. The area of interest is discretized into a regular grid of square cells (e.g., 25\,m side), with each cell centroid representing a node $v \in V$.

\subsubsection{Geospatial and Mobility Layers}
Each node is enriched with geospatial attributes derived from a \gls{dem} and national land-cover data, such as the Portuguese \gls{cos}~\cite{cos_dgt}. From these, elevation, slope, and terrain class are extracted. Edges $e=(u,v) \in E$ connect neighboring nodes if no static obstacle intersects the line segment between them. 

The temporal cost of traversing an edge is defined by Eq.~(\ref{eq:edge-time}), where $d(u,v)$ represents the geometric distance and $v_{\text{terrain}}$ the terrain-dependent speed:
\begin{equation}
t(u,v) = \frac{d(u,v)}{v_{\text{terrain}}(u,v)}.
\label{eq:edge-time}
\end{equation}
The term $v_{\text{terrain}}$ is computed using a configurable \gls{ugv} mobility model that maps land-cover and slope to expected vehicle speed. To ensure physical feasibility, steep slopes are penalized through the speed term, and continuous ascents beyond a cumulative gradient threshold are pruned during search.

\subsubsection{Military Intelligence as a Spatial Prior}
Rather than assuming deterministic threat positions, \gls{argus} incorporates uncertainty by representing each threat $i$ as a locational prior $L_i(r,c)$ over the discretized grid. This prior expresses the probability of threat presence at cell $(r,c)$, derived from terrain analysis and doctrinal heuristics. This probabilistic representation captures spatial uncertainty explicitly, providing a consistent foundation for subsequent risk computation.

\subsection{Probabilistic Risk Modeling}
Risk modeling in \gls{argus} follows the formulation $R = P \times I$, where $P$ is the probability of detection and $I$ the consequence of detection~\cite{nato2016ajp3}. This relationship is extended spatially to every cell of the operational grid.

\subsubsection{Parametric Detection Model}
The probability of detection as a function of distance is given by Eq.~(\ref{eq:pd_plateau_decay}), which defines a bounded, continuous curve consistent with radar detection theory:
\begin{equation}
P(d)=
\begin{cases}
1, & 0 \le d \le \phi R,\\[3pt]
\biggl(1-\Bigl(\dfrac{d-\phi R}{R-\phi R}\Bigr)^{2}\biggr)^{p}, & \phi R < d < R,\\[6pt]
0, & d \ge R.
\end{cases}
\label{eq:pd_plateau_decay}
\end{equation}
Here, $R$ is the effective detection range, $\phi$ defines the plateau fraction of certain detection, and $p$ controls the slope of decay. These parameters are provided by intelligence sources or derived from calibration data~\cite{Peto2013}. This function $P(d)$ is monotonic, bounded, and computationally efficient.

\subsubsection{Expected Detection Under Spatial Uncertainty}
Given a threat $i$ with locational prior $L_i(r,c)$, the expected probability of detection at a cell $(r,c)$ is computed using Eq.~(\ref{eq:expected_detection}) as a discrete spatial convolution:
\begin{equation}
\overline{P}_i(r,c) = 
\sum_{(r',c')} 
L_i(r',c') \cdot 
P_i(\|\mathbf{x}_{(r,c)} - \mathbf{x}_{(r',c')}\|_2),
\label{eq:expected_detection}
\end{equation}
where $\mathbf{x}_{(r,c)}$ represents the cell coordinates. Assuming independence among threats $\mathcal{T}$, the combined detection probability is obtained via survival composition in Eq.~(\ref{eq:multi_threat_pd}):
\begin{equation}
P_{\mathrm{det}}(r,c) = 
1 - \prod_{i \in \mathcal{T}} 
\bigl(1 - \overline{P}_i(r,c)\bigr).
\label{eq:multi_threat_pd}
\end{equation}
This formulation allows \gls{argus} to model overlapping detection regions in a computationally tractable manner.

\subsubsection{Operational Risk and Formation Footprint}
The per-cell operational risk is then computed as in Eq.~(\ref{eq:risk_cell}), combining detection probability and local impact:
\begin{equation}
R(r,c) = P_{\mathrm{det}}(r,c) \times I(r,c),
\label{eq:risk_cell}
\end{equation}
where $I(r,c)$ denotes the local consequence factor, which can encode threat lethality or mission sensitivity.  
To account for formation width, \gls{argus} applies the worst-case aggregation defined in Eq.~(\ref{eq:formation-worstcase}):
\begin{equation}
R_{\mathrm{form}}(r,c) = 
\max_{(u,v)\in\mathcal{N}_\rho(r,c)} R(u,v),
\label{eq:formation-worstcase}
\end{equation}
where $\mathcal{N}_\rho(r,c)$ defines a neighborhood whose radius $\rho$ depends on the commander-specified formation width $W_{\mathrm{form}}$. This conservative operator, ensures that all units within the formation corridor remain below acceptable exposure thresholds.

\subsubsection{Logarithmic Cost Transformation}
The cumulative survival probability of a path $\pi$ is given in Eq.~(\ref{eq:survival_product}), where each cell contributes independently:
\begin{equation}
P_S(\pi) = 
\prod_{(r,c)\in\pi} 
\bigl(1 - R_{\mathrm{form}}(r,c)\bigr).
\label{eq:survival_product}
\end{equation}
As multiplicative objectives are incompatible with additive-cost search algorithms such as \gls{astar}, \gls{argus} applies a logarithmic transformation, as shown in Eq.~(\ref{eq:log_risk}):
\begin{equation}
\ell(r,c) = -\log\bigl(1 - R_{\mathrm{form}}(r,c)\bigr),
\label{eq:log_risk}
\end{equation}
which yields an additive representation of risk.  
The accumulated log-risk along a path $\pi$ is then expressed by Eq.~(\ref{eq:path_logrisk}):
\begin{equation}
L(\pi) = \sum_{(r,c)\in\pi} \ell(r,c),
\label{eq:path_logrisk}
\end{equation}
allowing reformulation of the survival maximization problem into an equivalent minimization form (Eq.~\ref{eq:equivalence}):
\begin{equation}
\min_{\pi} L(\pi)
\quad \Leftrightarrow \quad
\max_{\pi} P_S(\pi).
\label{eq:equivalence}
\end{equation}
This equivalence preserves probabilistic meaning while enabling efficient graph-based optimization.

\subsection{Mission-Driven Planning Modes}
The resulting multi-cost graph serves as the foundation for mission optimization. The \gls{argus} framework operationalizes the commander’s intent through three distinct planning modes.

\begin{itemize}
    \item \textit{Balanced Mode:} minimizes a convex combination of normalized travel time and log-risk, as defined in Eq.~(\ref{eq:balanced_mode}):
    \begin{equation}
    \min_{\pi} \bigl(\alpha \, \hat{T}(\pi) + (1-\alpha)\, \hat{L}(\pi)\bigr),
    \label{eq:balanced_mode}
    \end{equation}
    where $\hat{T}(\pi)$ and $\hat{L}(\pi)$ are normalized values, and $\alpha \in [0,1]$ defines the trade-off between speed and survivability.
    
    \item \textit{Fast-within-Risk Mode:} minimizes total travel time under a per-cell risk constraint (Eq.~\ref{eq:fast_within_risk}):
    \begin{equation}
    \min_{\pi} T(\pi)
    \quad \text{s.t.} \quad 
    R_{\mathrm{form}}(r,c) \leq \bar{R}_{\max} 
    \;\; \forall (r,c)\in\pi.
    \label{eq:fast_within_risk}
    \end{equation}
    
    \item \textit{Safe-within-Time Mode:} minimizes cumulative log-risk subject to a total mission duration constraint (Eq.~\ref{eq:safe_within_time}):
    \begin{equation}
    \min_{\pi} L(\pi)
    \quad \text{s.t.} \quad 
    T(\pi) \leq T_{\max}.
    \label{eq:safe_within_time}
    \end{equation}
\end{itemize}

The final formulation in Eq.~(\ref{eq:safe_within_time}) corresponds to a \gls{rcspp} and represents the most computationally demanding case. The specialized APULSE algorithm, detailed in the next section, was developed to efficiently solve this constrained optimization problem for large-scale tactical scenarios.

\subsection{Local Repair Mechanism}\label{sec:method-dynamic-local}
The \gls{argus} framework includes a local repair mechanism that enables rapid adaptation when new threats or obstacles arise during mission execution. Instead of recomputing the entire path, only the affected segment is recalculated within a bounded patch window whose radius is defined by the threat range, half the formation width, and a configurable safety margin.  

The commander specifies a temporal slack parameter that controls how pronounced the correction can be: larger margins permit safer but longer detours, while smaller ones enforce stricter adherence to the original schedule. Within this corridor, three algorithms,\gls{apulse}, \gls{dstar}, and \gls{lpa} independently solve the \gls{rcspp} under the defined slack, each seeking to minimise cumulative log-risk. The framework automatically selects the path with the lowest overall risk, ensuring consistency with the global plan and achieving substantial computational savings compared with a full replan.

\section{The APULSE Algorithm for Time-Constrained Planning}
\noindent The \textit{Safe-within-Time} planning mode, which seeks the path of minimum risk subject to a mission time budget, constitutes the most computationally demanding task in the \gls{argus} framework. This problem can be formally defined as a \gls{rcspp}. Initial experiments with traditional multi-objective algorithms (e.g., NAMOA*), which explicitly compute the Pareto frontier between time and risk, revealed significant performance limitations. On the large-scale graph used in this work, these methods exhibited prohibitive computational overhead and failed to produce solutions within an operational timeframe—a behavior consistent with existing literature~\cite{Coego2012}.  

This challenge motivated the development of a specialized, hybrid algorithm designed to balance near-optimality with high performance on large, risk-aware graphs. This algorithm is referred to as APULSE.

\subsection{Problem Formulation}
Let $G=(V,E)$ be the planning graph, where each node $v\in V$ has a log-risk cost $\ell(v)$ and each edge $(u,v)\in E$ has a travel time $t(u,v)$.  
For a given start node $s$, goal node $g$, and maximum mission duration $B>0$, the optimization problem is defined in Eq.~(\ref{eq:rcspp}):
\begin{equation}
\min_{\pi\in\mathcal{P}(s,g)} 
\sum_{v\in\pi\setminus\{s\}} \ell(v)
\quad
\text{s.t.}\quad
\sum_{(u,v)\in\pi} t(u,v)\le B,
\label{eq:rcspp}
\end{equation}
where $\mathcal{P}(s,g)$ is the set of all simple paths connecting $s$ and $g$.  
The first term accumulates the logarithmic risk along the path, while the second imposes a hard upper bound on the total travel time. This formulation defines a \gls{rcspp} in which risk is minimized and time acts as a constrained resource.

\subsection{Algorithmic Design}
\begin{figure}[!t]
    \centering
    \includegraphics[width=\columnwidth]{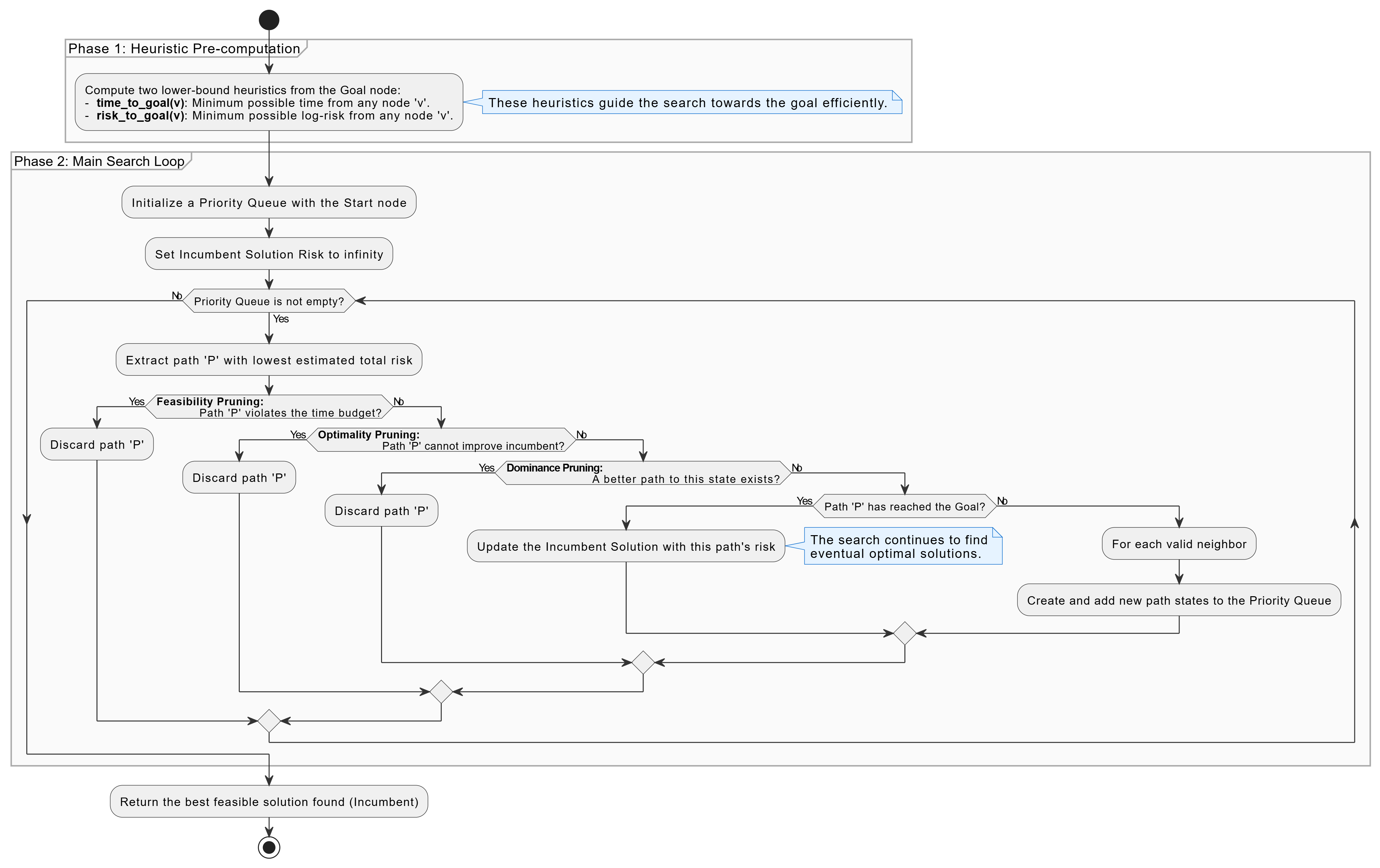}
    \caption{Workflow of the APULSE algorithm, showing the two main phases: heuristic pre-computation and the guided search loop with multi-stage pruning mechanisms.}
    \label{fig:apulse-workflow}
\end{figure}

The APULSE algorithm embodies the modern paradigm for solving the \gls{rcspp} on large networks by integrating three mechanisms: (i) \gls{astar}-style heuristic guidance, (ii) Pulse-inspired aggressive pruning, and (iii) lightweight state-space reduction.  
Its operational flow, illustrated in Fig.~\ref{fig:apulse-workflow}, is divided into two main phases: a pre-computation stage to establish heuristic foresight, and a guided search phase where candidate paths are explored and pruned.

\subsubsection{Phase 1: Heuristic Pre-computation}
Before the main search begins, APULSE computes two reverse single-source shortest-path maps from the goal node $g$:  
(1) the minimum time-to-goal, which provides an admissible lower bound on travel time from any node $v$ to $g$; and  
(2) the minimum log-risk-to-goal, which supplies a lower bound on cumulative log-risk.  
These heuristics supply the essential foresight for goal-directed exploration and enable powerful pruning in the subsequent phase.

\subsubsection{Phase 2: Guided Search with Multi-Stage Pruning}
The main loop of APULSE performs an informed best-first search.  
Let $g_\ell(v)$ and $g_t(v)$ denote, respectively, the accumulated log-risk and travel time from the start node $s$ to node $v$.  
Each partial path is prioritized according to the \gls{astar}-style evaluation function defined in Eq.~(\ref{eq:apulse-priority}):
\begin{equation}
f(v)=g_\ell(v)+\texttt{risk\_to\_goal}(v),
\label{eq:apulse-priority}
\end{equation}
which estimates the total log-risk of reaching the goal through $v$.  
Paths are extracted from a priority queue in ascending order of $f(v)$ and then subjected to three pruning tests before expansion.

\paragraph{Feasibility Pruning}
The first check enforces the time constraint.  
If the accumulated time $g_t(v)$ plus the minimum possible time-to-goal exceeds the mission budget $B$, the path is infeasible and discarded as expressed in Eq.~(\ref{eq:apulse-feasibility}):
\begin{equation}
g_t(v)+\texttt{time\_to\_goal}(v)>B \quad\Rightarrow\quad \text{discard}.
\label{eq:apulse-feasibility}
\end{equation}
This pruning stage prevents exploration of paths that cannot satisfy the operational deadline.

\paragraph{Optimality Pruning}
The second criterion removes paths whose estimated total risk is already worse than the current best solution.  
If $f(v)$ is greater than or equal to the incumbent log-risk $g_\ell^\star$, the path cannot yield improvement and is eliminated, as defined in Eq.~(\ref{eq:apulse-incumbent}):
\begin{equation}
f(v)\ge g_\ell^\star \quad\Rightarrow\quad \text{discard}.
\label{eq:apulse-incumbent}
\end{equation}

\begin{figure}[!t]
    \centering
    \includegraphics[width=0.8\columnwidth]{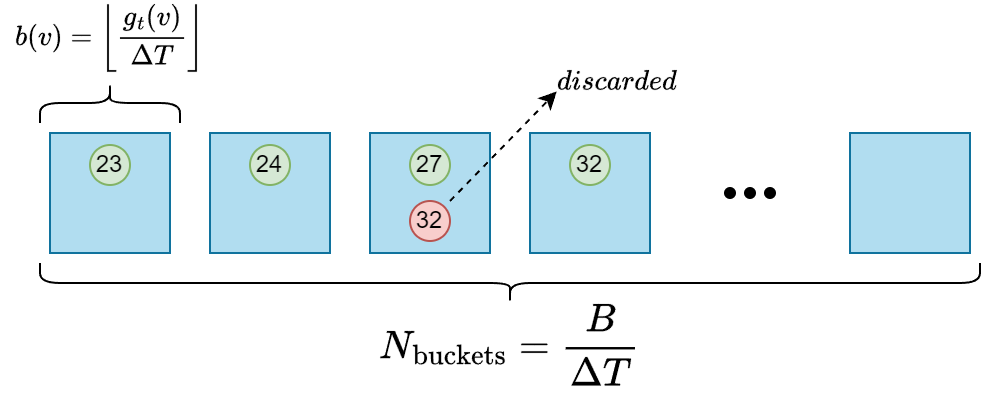}
    \caption{Illustration of the time-bucketing mechanism. The continuous time axis $g_t(v)$ is divided into discrete intervals of width $\Delta T$. For each state $(v,b)$ only the path with minimum cumulative log-risk $g_\ell^{\text{best}}(v,b)$ is stored, pruning dominated alternatives.}
    \label{fig:time-buckets}
\end{figure}

\paragraph{Dominance Pruning via Time Bucketing}
Finally, APULSE mitigates label proliferation through a state-space reduction method called time bucketing, illustrated in Fig.~\ref{fig:time-buckets}.  
The continuous time axis is discretized into intervals, or \textit{buckets}, of width $\Delta T$.  
When a path reaches node $v$ with accumulated time $g_t(v)$, it is assigned to a bucket $b(v)$ as defined in Eq.~(\ref{eq:apulse-bucket}):
\begin{equation}
b(v)=\left\lfloor\frac{g_t(v)}{\Delta T}\right\rfloor.
\label{eq:apulse-bucket}
\end{equation}
The pair $(v,b)$ defines a unique state.  
For each state, the algorithm retains only the path with the minimum cumulative log-risk.  
Any subsequent path arriving at the same node and bucket but with higher or equal risk is considered dominated and pruned.  
This mechanism drastically limits memory growth while preserving near-optimal solutions.

The bucket size $\Delta T$ defines a trade-off between precision and performance.  
APULSE employs an auto-tuning heuristic that sets $\Delta T\approx B/N$, where $B$ is the total mission budget and $N$ is a target number of buckets (empirically $8192$), ensuring balanced performance without manual calibration.

\subsubsection{Search Termination and Solution Update}
Once a path passes all pruning stages, its neighboring nodes generate new candidate paths added to the priority queue.  
When the goal node is reached, the incumbent solution is updated if the new path improves the cumulative log-risk.  
The process terminates when the queue becomes empty, guaranteeing that the returned path is the best feasible solution within the mission’s time constraint.  
This hybrid design enables near-optimal routing under stringent computational limits, providing operationally viable results for real-time mission planning in large-scale tactical environments.

\section{Implementation and Operational Demonstration}
\noindent To validate the proposed methodologies, the \gls{argus} framework was implemented as a fully functional software prototype. This section details the development environment, the operator-centric workflow, the suite of decision-support outputs, and the successful operational demonstration of the system's interoperability.

\subsection{Prototype Implementation and Workflow}
The \gls{argus} prototype was developed in Python 3.11, leveraging a suite of established scientific and geospatial libraries. Geospatial data processing was handled by \texttt{GeoPandas} and \texttt{Shapely}, graph construction and algorithmic execution by \texttt{NetworkX}, and numerical operations by \texttt{NumPy} and \texttt{SciPy}. The interactive user interface was built using \texttt{Streamlit}, which enabled rapid development of a mission-centric workspace.

The interface is structured around a clear, left-to-right narrative designed for intuitive operator use. A Control Panel on the left logically groups all inputs: drag-and-drop loaders for mission files (terrain grid, obstacles, threat data), a single selector for the optimization mode (which dynamically reveals relevant parameters like a time budget or risk ceiling), and sliders for formation settings. A key feature is the Tactical Mission Editor, an interactive map where operators can visually place start/end markers and, if necessary, manually define threat location priors. Upon execution, the results populate a Main Canvas on the right, which features a prominent banner with key performance indicators \glspl{kpi}---such as total distance, estimated time, and survival probability---and a tabbed structure for in-depth analysis.

\subsection{Decision Support and Visualization}
\begin{figure}[!t]
    \centering
    \subfloat[Operational Risk Map]{\includegraphics[width=0.48\columnwidth]{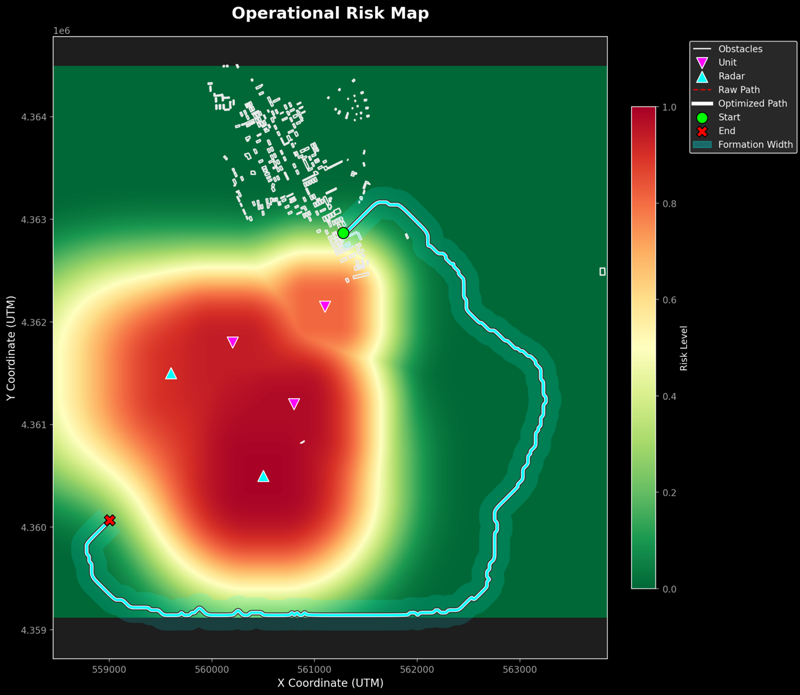}%
    \label{fig:argus_map_risk}}
    \hfill
    \subfloat[Land-Cover Map]{\includegraphics[width=0.48\columnwidth]{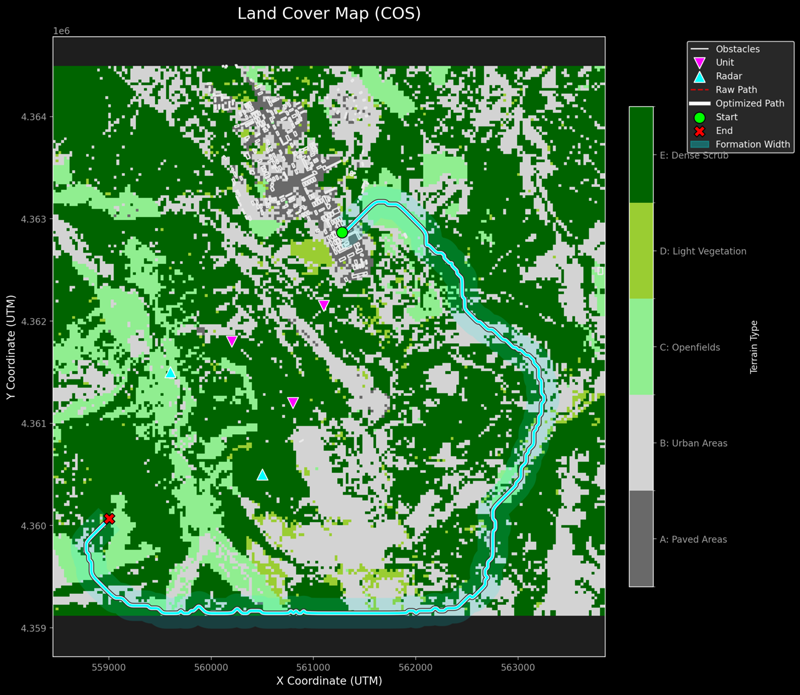}%
    \label{fig:argus_map_cos}}
    \caption{Primary 2D visualization outputs. The risk map (a) explains the survivability trade-offs of the computed path, while the land-cover map (b) provides context for its mobility and feasibility.}
    \label{fig:argus_mission_overview}
\end{figure}

\gls{argus} generates a comprehensive suite of visual and analytical outputs designed not just to present a path, but to provide explanatory evidence for its selection. The primary output is a pair of complementary 2D maps (Fig. \ref{fig:argus_mission_overview}): an Operational Risk Map, which renders the risk field as a continuous heat surface, and a Land-Cover Map, which provides context for mobility and time estimates.

Deeper analysis is supported by an analytical dashboard that includes a longitudinal altitude profile of the route, a quantitative breakdown of risk exposure (e.g., distance traveled in high-risk zones), an analysis of terrain composition, and a table reporting the \gls{cpa} to each known threat. This multi-faceted presentation ensures the commander has full situational awareness and can validate that the proposed course of action aligns with mission requirements.

\subsection{Operational Demonstration in a Field Exercise}
\begin{figure}[!t]
    \centering
    \subfloat[The Vigilant UGV platform used during ARTEX 25.]{\includegraphics[width=0.45\columnwidth]{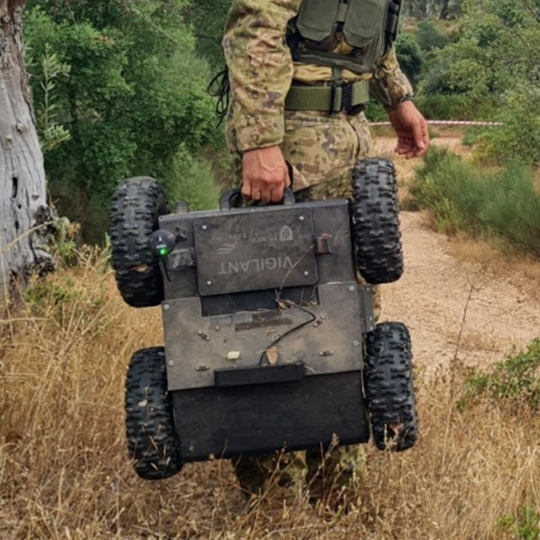}%
    \label{fig:ugv_vigilant}}
    \hfill
    \subfloat[ARGUS-generated path imported into QGroundControl.]{\includegraphics[width=0.52\columnwidth]{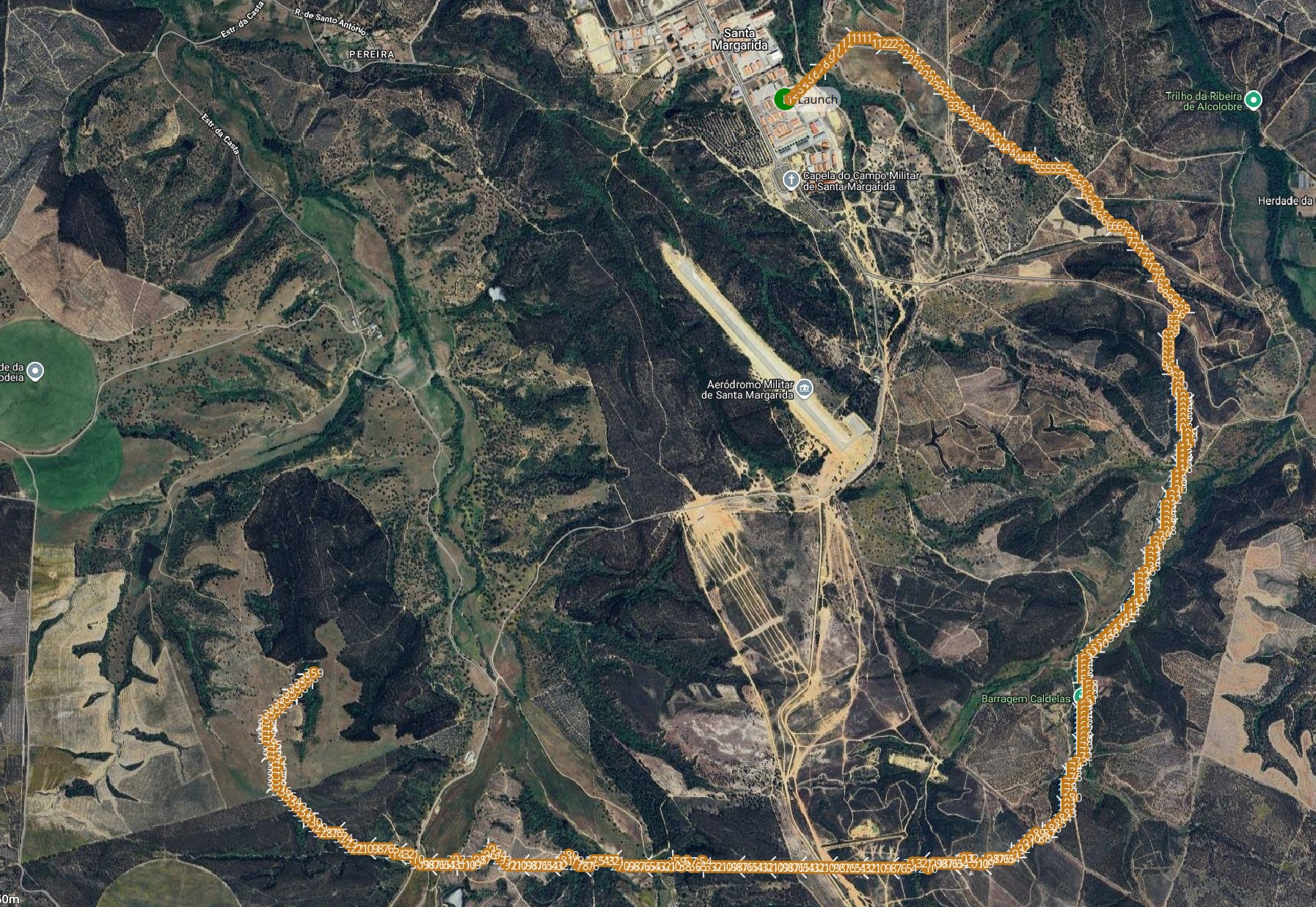}%
    \label{fig:qground_2}}
    \caption{Operational demonstration of ARGUS interoperability.}
    \label{fig:artex_demo}
\end{figure}

A critical validation of the framework was conducted during the Portuguese Army's \gls{artex} 25 field exercise. The objective was to demonstrate the end-to-end interoperability of \gls{argus} with a real mission-control platform. The workflow was tested using the Vigilant \gls{ugv} (Fig. \ref{fig:artex_demo}a), a remotely operated surveillance and reconnaissance platform.

Once a path was computed and validated within the \gls{argus} interface, it was exported in the standard \texttt{.waypoints} format. This file was then imported into the QGroundControl mission-control software (Fig. \ref{fig:artex_demo}b), the same application used for field control of the \gls{ugv}. This test validated the entire planning-to-execution pipeline, confirming that \gls{argus} can generate tactically routes that are directly consumable by operational systems. A key insight from this experiment was the dual utility of the formation corridor model; beyond accounting for physical footprint, it provides operators with a pre-vetted maneuvering margin for minor, real-time navigational adjustments.

\section{Experimental Validation}
\noindent Having demonstrated the operational workflow in the previous section, this part presents the experimental validation of both the \gls{argus} framework and the underlying APULSE algorithm. The evaluation was designed to assess two complementary dimensions: (i) mission-level effectiveness in producing tactically sound paths and (ii) algorithmic efficiency and scalability under time-constrained conditions. All experiments were executed on an Intel i9 Ultra 32 GB RAM laptop.

\subsection{ARGUS Framework: Mission-Level Evaluation}
This first part of the evaluation assesses the framework's primary function: translating a commander's intent into strategically sound and distinct courses of action. The focus is on the quality of the outputs—such as survival probability and mission time—rather than on computational performance and the tests were conducted in a 30 km$^{2}$ region of the Campo Militar de Santa Margarida represented by over 46 000 nodes.

\subsubsection{Analysis of Mission Mode Flexibility}
\begin{figure}[!t]
    \centering
    \subfloat[$\alpha = 0.0$ (Safest Path)]{%
        \includegraphics[width=0.48\columnwidth]{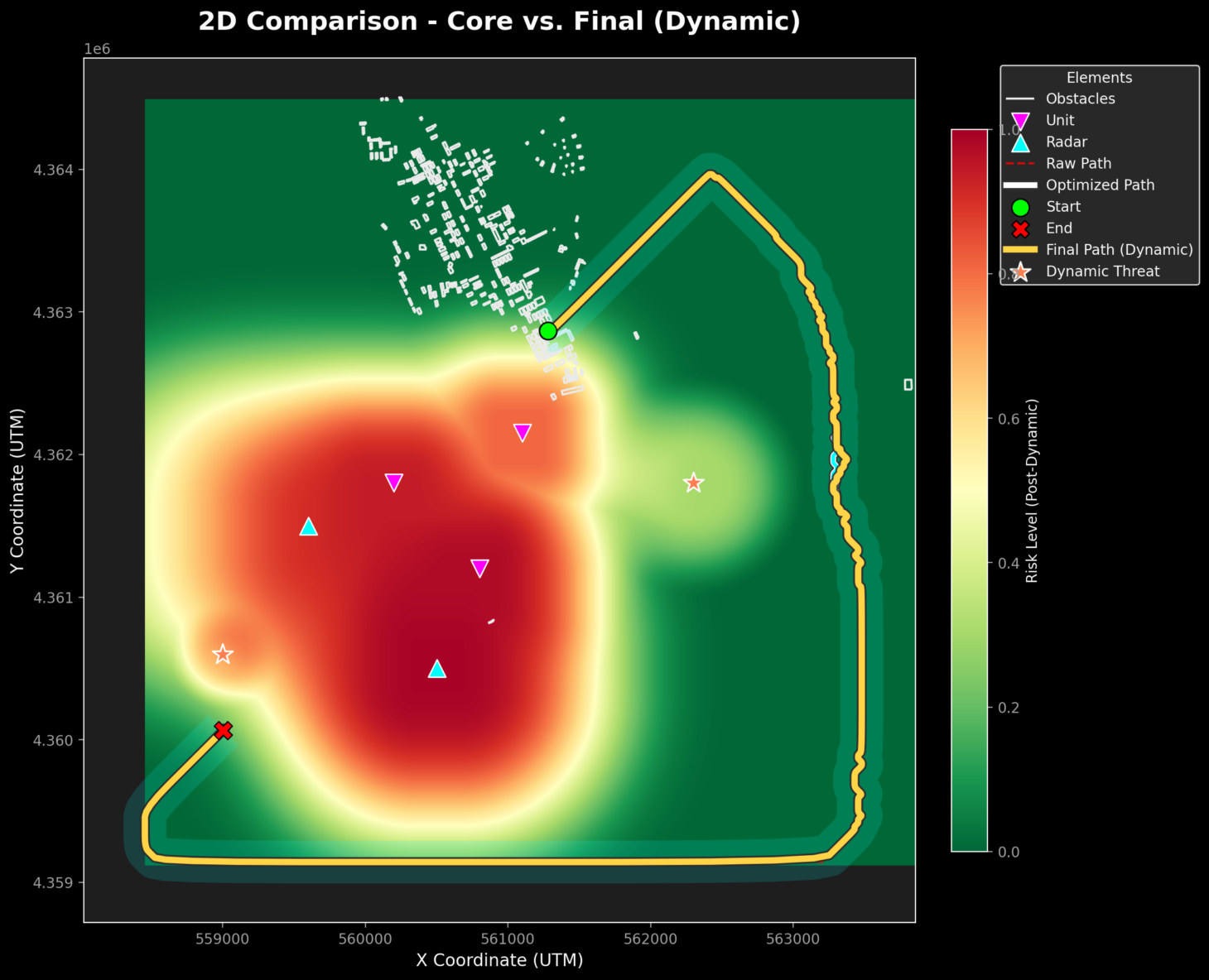}%
        \label{fig:alpha00}}
    \hfill
    \subfloat[$\alpha = 1.0$ (Fastest Path)]{%
        \includegraphics[width=0.48\columnwidth]{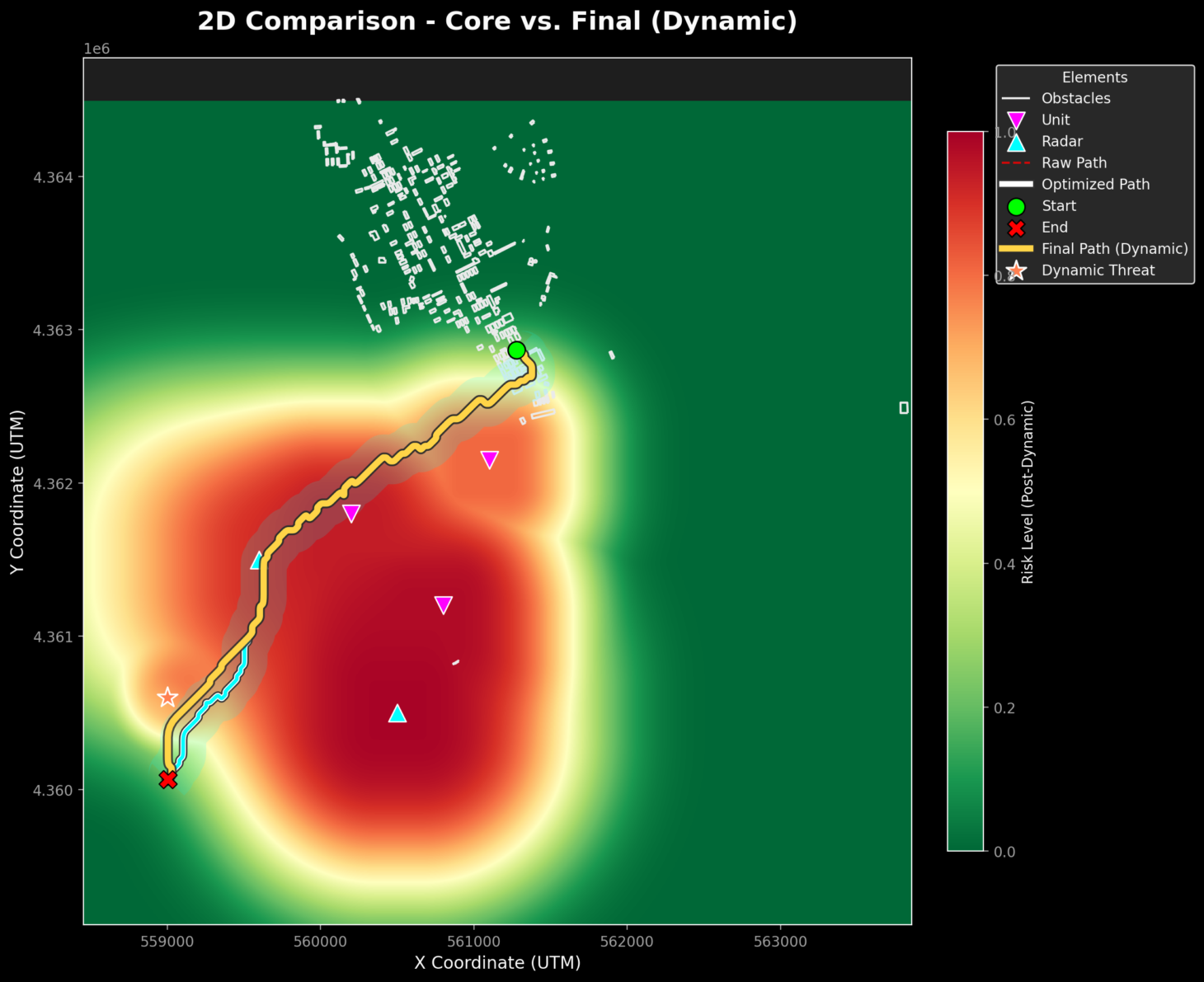}%
        \label{fig:alpha10}}
    \caption{Comparison of paths generated by the Balanced Mode for extreme weighting parameters $\alpha$, illustrating the trade-off between mission duration and survivability.}
    \label{fig:balanced_paths}
\end{figure}

To demonstrate the framework's ability to adapt to varying tactical priorities, a consistent long-distance planning scenario was executed across all seven configurations of the three main operational modes. A constant formation width of 300 m was applied to ensure a consistent worst-case risk assessment. The results, summarized in Table~\ref{tab:mode-comparison} and some of them illustrated in Fig.~\ref{fig:balanced_paths}, highlight the strategic trade-offs generated by each mode.

In the \textit{Balanced Mode}, shown in Fig.~\ref{fig:balanced_paths}, the two extreme configurations of the weighting parameter $\alpha$ clearly illustrate the operational spectrum. The safest path ($\alpha=0.0$, Fig.~\ref{fig:alpha00}) is a 13.17 km circuitous route that follows the perimeter of the operational area to bypass central high-risk zones, requiring approximately 132 min. In contrast, the fastest path ($\alpha=1.0$, Fig.~\ref{fig:alpha10}) is a direct 4.45 km route completed in under 20 min but with near-zero survival probability.

Similarly, the \textit{Safe-within-Time Mode} demonstrates how the APULSE algorithm effectively exploits available temporal slack to enhance survivability. When the mission budget increased from 30 min to 40 min, the algorithm found a longer and more evasive route that increased survival probability from 2.29 \% to 12.06 \%—a fivefold improvement. The results for the \textit{Fast-within-Risk Mode} also confirmed its operational robustness, including the correct activation of its fallback mechanism when an overly strict risk ceiling rendered a direct path infeasible. These findings confirm that the \gls{argus} framework successfully translates high-level strategic priorities into tangible, diverse, and tactically coherent courses of action.

\begin{table}[!t]
\centering
\caption{Key performance indicators for mission-mode simulations.}
\label{tab:mode-comparison}
\resizebox{\columnwidth}{!}{%
\begin{tabular}{llccc}
\toprule
Mode & Parameter & Distance (km) & Time (min:s) & Survival Prob. \\
\midrule
\multirow{2}{*}{Balanced} 
& $\alpha=0.0$ (Safest)  & 13.17 & 132:41 & 36.27 \% \\
& $\alpha=1.0$ (Fastest) & 4.45  & 19:44  & 0.00 \%  \\
\midrule
\multirow{2}{*}{Fast-within-Risk} 
& Max Risk = 0.15 & 13.17 & 132:41 & 36.27 \% \\
& Max Risk = 0.50 & 6.62  & 24:03  & 0.04 \%  \\
\midrule
\multirow{2}{*}{Safe-within-Time} 
& Budget = 30 min & 6.62  & 29:59  & 2.29 \%  \\
& Budget = 40 min & 8.64  & 39:59  & 12.06 \% \\
\bottomrule
\end{tabular}%
}
\end{table}

\subsubsection{Effectiveness of the Dynamic Replanning Module}
A defining capability of the \gls{argus} framework is its ability to react to dynamic events—such as the sudden appearance of new threats—by locally repairing an existing route instead of recalculating it from scratch. This experiment evaluated the effectiveness of that dynamic re-planning mechanism under realistic operational conditions.

\begin{itemize}
    \item \textbf{Dynamic Event Scenario:}
    Each of the seven baseline simulations described previously was re-executed under identical conditions, but with a set of new threats introduced midway along the path. These threats shared identical characteristics and spatial priors across all tests, ensuring consistent evaluation. The initial planned route was thus rendered unsafe, compelling the system to perform a localized recalculation (or “patch”) to bypass the new danger zones. This approach simulates an evolving battlefield, where situational awareness updates force the command system to adapt without compromising mission continuity.

    \item \textbf{Patching and Rerouting Results:}
    The comparative results of this process are summarized in Table~\ref{tab:dynamic-results}. The table reports pre- and post-patch metrics for each mission mode, including log-risk, traversal time, and survival probability, together with the relative or absolute variation (\(\Delta\)) observed after the re-plan. 

    \item \textbf{Discussion and Tactical Interpretation:}
    The data presented in Table~\ref{tab:dynamic-results} confirm that the dynamic re-planning mechanism is critical for maintaining mission viability in evolving threat environments. This recalculation process highlights two primary, parallel outcomes. \textbf{First}, the patching algorithm consistently identified alternate routes that substantially improved survivability; in configurations such as the \textit{Safe-within-Time} mode, this achieved up to 30\% reductions in log-risk and increases in survival probability. \textbf{Second}, this improvement naturally entails additional travel time, representing the tactical trade-off between risk mitigation and temporal efficiency; in those same missions, local patching extended total duration by up to 50\% while reducing exposure by one-third.

    The \gls{argus} framework mitigates this tension through a commander-configurable temporal slack parameter that governs the allowable deviation from the nominal time budget during re-planning. This feature preserves human decision authority while ensuring the \gls{ugv}'s autonomy remains adaptive to emerging intelligence inputs.
\end{itemize}

\begin{table}[!t]
\centering
\caption{Comparison of pre- and post-patch metrics across all dynamic re-planning scenarios.}
\label{tab:dynamic-results}
\resizebox{\columnwidth}{!}{%
\begin{tabular}{lccrcrcr}
\toprule
& & \multicolumn{2}{c}{Risk (log)} & \multicolumn{2}{c}{Time (s)} & \multicolumn{2}{c}{Survival Prob.} \\
\cmidrule(lr){3-4}\cmidrule(lr){5-6}\cmidrule(lr){7-8}
Mode & Parameter & Pre & Post ($\Delta$) & Pre & Post ($\Delta$) & Pre & Post ($\Delta$) \\
\midrule
\multirow{2}{*}{Balanced} 
& $\alpha=0.0$ & 20.28 & 20.28 (0.0\%) & 7962 & 7870 (−1.2\%) & 36.3\% & 36.3\% (+0.0) \\
& $\alpha=1.0$ & 246.66 & 240.54 (−2.5\%) & 1184 & 1425 (+20.4\%) & 0.0\% & 0.0\% (+0.0) \\
\midrule
\multirow{2}{*}{Fast-within-Risk} 
& Max Risk = 0.15 & 20.28 & 20.28 (0.0\%) & 7962 & 7870 (−1.2\%) & 36.3\% & 36.3\% (+0.0) \\
& Max Risk = 0.50 & 159.62 & 151.87 (−4.9\%) & 1444 & 1718 (+19.0\%) & 0.03\% & 0.05\% (+0.02) \\
\midrule
\multirow{2}{*}{Safe-within-Time} 
& Budget = 30 min & 84.34 & 58.77 (−30.3\%) & 1800 & 2697 (+49.8\%) & 1.5\% & 5.3\% (+3.8) \\
& Budget = 40 min & 61.01 & 45.80 (−24.9\%) & 2400 & 2999 (+25.0\%) & 4.7\% & 10.1\% (+5.4) \\
\bottomrule
\end{tabular}
}
\end{table}

\subsection{APULSE Algorithm: Performance Benchmark}
\noindent A comprehensive benchmark was conducted to evaluate the computational performance and scalability of the hybrid \gls{apulse} algorithm. The comparison included three state-of-the-art reference algorithms from the public repository of Ahmadi~\textit{et~al.}~\cite{Ahmadi2024EnhWCSP}: the unidirectional \gls{wc-a} and two bidirectional variants, \gls{wc-ba} and \gls{wc-ebba}. All algorithms were executed under identical hardware and graph conditions derived from the same \gls{argus} terrain model of the Santa Margarida military field.

Four test instances of increasing spatial separation between start and goal nodes were defined: Small (1$\to$1k), Medium (1$\to$15k), Med-Large (1k$\to$25k), and Large (5k$\to$46k). For each instance, the time budget $B$ was systematically varied using a slack parameter $\alpha$, representing the additional temporal allowance over the minimum feasible path time $T_{\min}$. Tight budgets (low $\alpha$) favor aggressive feasibility pruning and shorter runtimes, whereas loose budgets (high $\alpha$) enlarge the search space, stressing algorithm scalability. Each configuration was executed multiple times, with runs exceeding a ten-minute timeout marked as failures.

\subsubsection{Results and Analysis}
The quantitative outcomes of the benchmark are summarized in Table~\ref{tab:rcspp-main}, which reports average runtimes (in seconds) across the four instance scales for $\alpha \in \{0.10, 0.20, 0.50\}$. Failures or runs exceeding the timeout are denoted by “—”.  

\begin{table}[!t]
\centering
\caption{Average runtime (s) across instance scales and main budgets.}
\label{tab:rcspp-main}
\resizebox{\columnwidth}{!}{%
\begin{tabular}{llrrr}
\toprule
Scale & Algorithm & $\alpha=0.10$ & $\alpha=0.20$ & $\alpha=0.50$ \\
\midrule
\multirow{4}{*}{Small (1$\to$1k)}
& \gls{apulse} & 0.051 & 0.117 & 0.516 \\
& \gls{wc-a} & \textbf{0.001} & \textbf{0.001} & \textbf{0.000} \\
& \gls{wc-ba} & 0.002 & 0.002 & 0.002 \\
& \gls{wc-ebba} & 0.001 & 0.001 & 0.001 \\
\midrule
\multirow{4}{*}{Medium (1$\to$15k)}
& \gls{apulse} & \textbf{1.404} & \textbf{5.327} & 27.005 \\
& \gls{wc-a} & 49.931 & 49.415 & 50.250 \\
& \gls{wc-ba} & 5.567 & 5.847 & \textbf{11.790} \\
& \gls{wc-ebba} & 13.085 & 14.757 & 15.693 \\
\midrule
\multirow{4}{*}{Med-Large (1k$\to$25k)}
& \gls{apulse} & \textbf{1.538} & \textbf{4.737} & \textbf{21.060} \\
& \gls{wc-a} & 27.617 & 29.206 & 30.302 \\
& \gls{wc-ba} & 103.567 & 105.380 & 97.331 \\
& \gls{wc-ebba} & 123.548 & 119.190 & 122.836 \\
\midrule
\multirow{4}{*}{Large (5k$\to$46k)}
& \gls{apulse} & \textbf{1.191} & \textbf{5.873} & \textbf{21.602} \\
& \gls{wc-a} & 60.047 & 45.447 & 49.818 \\
& \gls{wc-ba} & — & — & — \\
& \gls{wc-ebba} & — & — & — \\
\bottomrule
\end{tabular}
}
\end{table}

At small scales, the reference algorithms outperform due to negligible initialization overheads. However, as the problem size increases, the hybrid design of \gls{apulse} achieves better scalability. From the medium scale onward, it consistently yields the lowest runtimes, and for the largest instance, it remains the only algorithm completing all configurations within the operational timeout.  

\begin{figure}[!t]
    \centering
    \subfloat[Runtime vs. instance scale ($\alpha=0.5$)]{%
        \includegraphics[width=0.48\columnwidth]{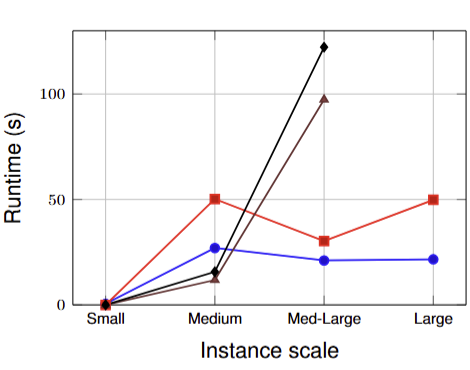}%
        \label{fig:scale-benchmark}}
    \hfill
    \subfloat[Runtime vs. slack $\alpha$ (medium scale)]{%
        \includegraphics[width=0.48\columnwidth]{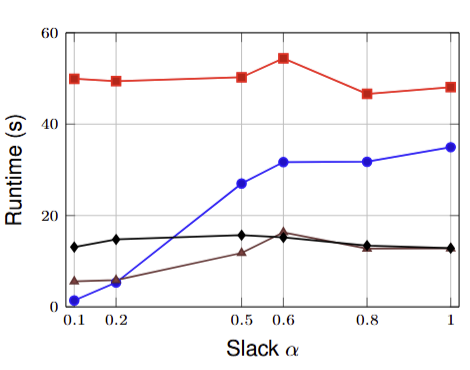}%
        \label{fig:slack-benchmark}}
    \caption{APULSE benchmark results. (a) Runtime scalability with increasing instance size. (b) Runtime behavior with varying budget slack. Missing points indicate timeout or failure.}
    \label{fig:apulse_benchmarks}
\end{figure}

The overall runtime behavior is summarized in Fig.~\ref{fig:apulse_benchmarks}.  
Fig.~\ref{fig:scale-benchmark} plots runtime against instance size for a moderate slack ($\alpha=0.5$), while Fig.~\ref{fig:slack-benchmark} shows runtime variation with increasing slack for a fixed medium instance. These plots clearly demonstrate that while the runtime of all solvers grows with problem size or relaxed constraints, \gls{apulse} maintains smooth scalability and robustness across both dimensions.

These results confirm that \gls{apulse} achieves both speed and robustness across a wide range of problem sizes and constraint levels. The bidirectional variants (\gls{wc-ba}, \gls{wc-ebba}) are competitive only under extremely tight constraints and become unstable as the feasible region expands, whereas \gls{apulse} sustains predictable performance throughout.

\subsubsection{Solution Quality}
Beyond computational efficiency, the quality of the solutions produced by \gls{apulse} was validated against the exact optimal paths.  
Across all 26 benchmark configurations represented in Table~\ref{tab:rcspp-main}, \gls{apulse} achieved the exact optimal solution in 25 cases.  
In the single suboptimal instance, corresponding to the Med-Large scale at $\alpha=0.50$, the deviation from the optimal risk was only 0.0025\%, a negligible trade-off attributable to temporal discretization in the time-bucketing mechanism.  

\begin{table}[!t]
\centering
\caption{Optimality of \gls{apulse} compared to the exact optimal path ($\rho^*$).}
\label{tab:rcspp-quality}
\resizebox{0.75\columnwidth}{!}{%
\begin{tabular}{lc}
\toprule
Optimality Outcome & Number of Instances ($N=26$) \\
\midrule
Optimal Path Achieved ($\rho_{\text{APULSE}} = \rho^*$) & 25 \\
Suboptimal Path Found (Deviation $\approx 0.0025\%$) & 1 \\
\bottomrule
\end{tabular}
}
\end{table}

This minor deviation demonstrates the efficiency–precision balance intrinsic to \gls{apulse}.  
The results validate that its hybrid heuristic–pruning architecture provides a near-optimal trade-off between solution accuracy and computational tractability, enabling real-time performance without compromising mission-level reliability.

\subsubsection{Summary}
In summary, the benchmark confirms the superior performance of the \gls{apulse} algorithm for solving the \gls{rcspp} within the \gls{argus} framework.  
It achieved exact optimality in 25 of 26 cases while outperforming all reference algorithms in runtime and scalability across all but the smallest instances.  
Unlike bidirectional approaches, \gls{apulse} maintained stable convergence across all tested configurations, never exceeding the ten-minute operational limit. These results confirm that \gls{apulse} provides a robust and computationally efficient solution for time-critical, large-scale mission planning under risk-aware constraints.

\section{Conclusion}
\noindent This work presented the \gls{argus} framework, a mission-oriented system that translates a commander's intent and operational intelligence into optimized, risk-aware paths for \glspl{ugv}. The main contributions span three levels: a unified probabilistic risk model integrating spatial intelligence and formation footprint; the hybrid \gls{apulse} algorithm for efficient \gls{rcspp} resolution; and a validated software prototype demonstrating full interoperability with real mission-control systems.

The study was bounded by simplifying assumptions, such as static terrain data and idealized sensor models, which define avenues for future development. Near-term work will refine the risk model to incorporate occlusion and dynamic factors, integrate real-time data streams, and extend the optimization framework with energy and vehicle constraints.  

In summary, this work provides a contribution to the emerging field of military operations involving \glspl{ugv}, advancing the integration of risk modeling into mission planning and delivering a framework that supports the tactical decision-making process.


\bibliographystyle{IEEEtran}
\bibliography{referencias}


\end{document}